# Experimental Characterization and Dynamic Modeling of THz Channels Under Fog Conditions

Jiaobiao Zhao[1], Kefeng Huang[1], Xiaoxiang Li[1], Mingxia Zhang[1], Peian Li[1], Jie Yang[1], Wenbo Liu[1], Yiming Zhao[1,2,3], Weidong Hu[1,3], Jianjun Ma[1,3]

[1]School of Integrated Circuits and Electronics, Beijing Institute of Technology, Beijing 100081, China

[2]School of Interdisciplinary Science, Beijing Institute of Technology, Beijing, 100081 China

[3]State Key Laboratory of Environment Characteristics and Effects for Near-space, Beijing 100081, China

Corresponding author: Jianjun Ma

**Abstract**—The terahertz (THz) band is a promising candidate for sixth-generation wireless networks, but its deploymen in outdoor environments is challenged by meteorological phenomena, particularly fog, which imposes variable and difficult-to-predict channel degradation. This article introduces dynamic channel model for the THz band explicitly driven by the time-evolving droplet size distribution (DSD) of fog, integrating real-time microphysical sensing to capture variations in the fog microstructure. Experimental measurements were conducted at 220 GHz and 320 GHz in a controlled fog chamber to achieve quasi-stationary states, and a larger room-scale setup to characterize dynamic, non-stationary fog evolution. The results confirm that channel power loss is overwhelmingly dominated by absorption rather than scattering, validating the use of the computationally efficient Rayleigh approximation below 1 THz. Statistical analysis revealed exceptionally high Rician K-factors, demonstrating that THz channels maintain strong line-of-sight stability even in dense fog. System-level performance analysis shows that degradation in bit error rate is driven by the slow, gradual evolution of the DSD, rather than fast multipath fading. This finding enables the reliable simplification of the THz fog channel into a near-Gaussian channel model with time-varying signal-to-noise ratio. This microphysics-aware approach established here provides the necessary foundation for developing adaptive system designs centered on SNR tracking for robust future THz networks

# I. Introduction

The relentless demand for higher data rates has propelled the terahertz (THz) band (0.1-10 THz) into prominence as a potential enabler for sixth-generation (6G) wireless communication and sensing networks [1, 2]. Owing to its vast spectral resources, THz channels promise ultra-broadband capabilities supporting data rates in the terabit-per-second (Tbps) regime [3, 4], essential for latency-sensitive and bandwidth-hungry applications, such as holographic telepresence, immersive extended reality, and massive machine-type communication. However, the real-world deployment of THz communication is fundamentally constrained by strong interactions with atmospheric media. [5]. Beyond free-space molecular absorption, notably from water vapor and oxygen [6], meteorological phenomena like rain, snow, and fog introduce significant, time-variable extinction losses [7].

Fog poses a unique challenge because its microphysical properties - specifically the droplet size distribution (DSD) - can evolve rapidly and non-linearly, making reliable prediction difficult. The most commonly adopted framework for modeling fog attenuation is the International Telecommunication Union Radiocommunication Sector (ITU-R) Recommendation P.840 [8]. This empirical model treats attenuation ($\gamma_g$) as a linear function of liquid water content ($\rho_l$), $\gamma_g = k_l \rho_l$, with the coefficient $k_l$ derived from the dielectric properties of water and the operating frequency. While useful for preliminary link budgets, this deterministic, LWC- (or visibility-) based approach fundamentally fails to account for the dynamic variability imposed by the evolution of the DSD.

Prior experimental work has provided valuable insights, showing, for instance, that broadband THz pulses can traverse dense fog (approx. 7 m visibility) with only modest additional power loss relative to clear weather, supporting the view that broadband THz channels could be promising for landings of airplanes under severe conditions of fog [9]. Similarly, a 330 GHz FMCW-radar chamber study revealed not only power loss but also incremental group delay, underscoring that droplet-induced channel degradation can perturb both communication and ranging [10]. Comparative measurements of 625 GHz and 1.55 μm infrared (IR) links further showed that while the IR channel experienced severe power loss and BER collapse, the THz link remained comparatively robust, highlighting its resilience under low-visibility fogy events [7].

Despite these insights, some modeling efforts have introduced stochastic approaches, using Gamma-distributed statistics to analyze capacity and outage to analyze signal-ro-noise ratio (SNR), outage, and capacity under α-μ fading [11]. But these still generally treat the fog medium as static with fixed microphysical parameters. Discrepancies between theoretical predictions and measured results persist largely due to the insufficient representation of DSD dynamics and, crucially, the lack of integrated, real-time microphysical sensing coupled directly with channel measurements. Recent THz channel sounding efforts emphasize the critical need for dynamic channel models that reveal a clear correspondence between the time-varying physical environment and the resulting channel characteristics.

In this work, we try to address these limitations by developing a microphysics-aware dynamic model for THz channels that is explicitly driven by the measured, evolving DSD of fog. A dynamic, microphysics-aware THz channel model is developed and experimentally validated, through dual-scale experimental characterizations (controlled chamber and large-scale room). The following of this article is organized as follows: Section II presents the experimental characterization of THz channel performance in a controlled fog chamber, including the experimental setup, theoretical models based on Rayleigh approximation and Mie theory, channel measurement and modeling results, and channel power profile analysis. Section III extends the investigation to a larger room-scale foggy environment, describing the experimental setup and presenting channel measurement and modeling results under more realistic, quasi-non-stationary fog conditions. Section IV analyzes the system-level bit error rate performance under fog, demonstrating the dominance of gradual DSD evolution over fast multipath fading. Finally, Section V concludes the article with a summary of key findings and suggestions for future work.

## II. Channel performance in fog chamber

**1. Channel sounding system architecture**

The experimental platform was designed to emulate a line-of-sight (LoS) THz channel transmitting through a controlled fog environment, as shown in Fig. 1(a). At the transmitter side, a Ceyear 1465D signal generator produces a baseband signal (100 kHz-20 GHz), which is upconverted to the target THz range using a Ceyear 82406D ×18 frequency multiplier. The radiated signal is launched by a Ceyear 89901S horn antenna integrated with a 10-cm dielectric lens, yielding a measured gain up to 33 dBi at 220 GHz and a beamwidth of ~4°. The receiver employs an identical horn-lens assembly, ensuring symmetric channel characteristics. A Ceyear 71718 power sensor records received power, with measurements sampled at 5 Hz sampling rate through a synchronized computer-controlled acquisition system. Both horn antennas are mounted 85 cm above the ground - well above the Fresnel zone radius (~ 3.08 cm) - to eliminate ground effects. A purpose-built chamber with inner dimensions of $1.0 \times 0.5 \times 0.5$ m$^3$ (in L×W×H) provides the propagation medium. Circular apertures of 90 mm diameter are machined into the chamber walls to permit unobstructed channel propagation through it. The entire system, including signal generation, frequency conversion, power detection, and environmental monitoring, is integrated for real-time operation with high mechanical stability to ensure reliable measurements.

Dense fog is introduced into the chamber using a commercial ultrasonic generator, which generates water droplets with diameters ranging from 1 to 30 μm. The fog generator output is fed into the chamber via flexible tubing, and airflow is carefully controlled to maintain quasi-stationary state of fog during experiments, which is necessary for robust small-scale statistical analysis (e.g., Rician fitting). Microphysical characterization is performed using a Furbs FBS-310B laser particle size analyzer operating at 1 Hz sampling rate. The analyzer, aligned on a rigid optical rail, projects a 635-nm laser beam that propagates through the fog, as shown in Fig 1(a). For practical reasons, the laser beam is placed adjacent to rather than collinear with the THz channel path, avoiding interference with the channel transmission. While this arrangement does not yield an exact sample of the channel volume, the co-located sampling height ensures that the retrieved DSD provides a reliable representation of the fog conditions experienced by the THz channel beam.

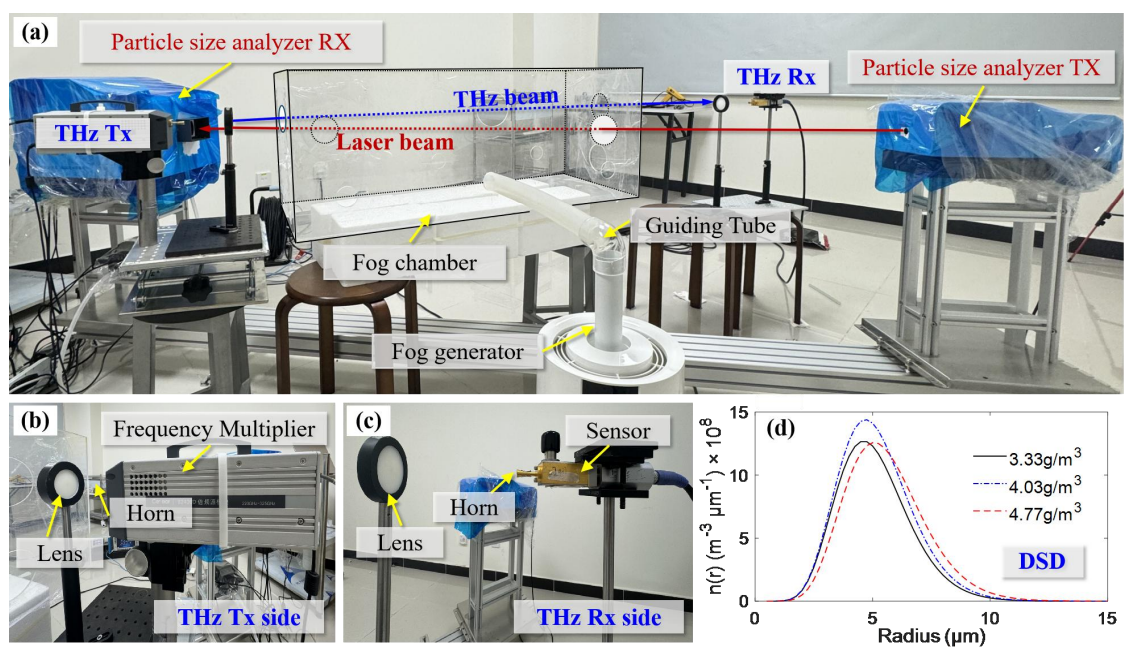

**Figure 1** (a) Experimental setup for channel measurement in fog chamber with (b) THz transmitter side, (c) THz receiver side and (d) variation of droplet size distribution (DSD) at three different times.

## 2. Theoretical models

The total power loss suffered by THz channels in fog arises from molecular absorption by atmospheric gases, primarily water vapor, and extinction caused by absorption and scattering from suspended fog droplets. The ITU-R Recommendation P.840 provides an empirical framework for estimating fog-induced attenuation as a linear function of the LWC [8], with frequency-dependent coefficients derived from the complex permittivity of water. While useful for link-budget estimates, this approach does not account for the microphysical variability of fog. For a more accurate description, scattering theory is required. Mie theory rigorously models the extinction cross section when droplet radius are comparable to the incident wavelength [12, 13]. In the case of fog, where droplet radius typically range from 1-25 μm [14, 15], the size parameter $\chi = \pi r/\lambda$ is much smaller than unity at frequencies below ~1 THz, and the Mie solution converges to the Rayleigh approximation [16]. In this regime, power loss is dominated by absorption within the droplets, while scattering contributions remain negligible.

Since extinction depends explicitly on the droplet size distribution (DSD), statistical characterization of fog microphysics is required. A modified gamma distribution is commonly used to represent DSDs [17], as $n(r) = N_0 r^\alpha \exp(-br^\beta)$, where $n(r)$ denotes the number density of droplets with radius between $r$ and $r + \mathrm{d}r$, $N_0$ controls the total droplet concentration, and $\alpha, \beta, b$ are empirical shape and scale parameters. Measurements from our fog chamber confirm that the DSD is time-varying (see Fig. 1 (d)), even under identical generation conditions, evolving on time scales of seconds as the fog field develops and dissipates. This observation is consistent with prior findings for rainfall microphysics [5]. Because extinction efficiency in the Rayleigh regime increases rapidly with droplet radius, even moderate shifts toward larger fog droplets or broader distribution tails lead to significant increases in THz channel power loss. Consequently, channel degradation cannot be reliably represented by a single LWC or visibility value, but must instead be modeled as a dynamic function of DSD, $n(r, t)$, as

$$n(r,t) = N_0(t) r^{\alpha(t)} \exp[-b(t) r^{\beta(t)}]. \qquad (1)$$

Formally, the attenuation coefficient at time $t$ is express as

$$\gamma_f(t) = 3.3429 \int_0^\infty \xi_e(r, \lambda) \cdot \pi r^2 \cdot n(r,t) dr \qquad (2)$$

where $\xi_e(r, \lambda)$ is the extinction efficiency in the Rayleigh approximation [18], and can be expressed as

$$\xi_e(r, \lambda) = 4\chi \,\mathrm{Im}\{-K\} + \frac{8}{3}\chi^4 |K|^2 + \cdots \qquad (3)$$

where $K$ is a complex quantity defined in terms of the complex dielectric property $\varepsilon$ of the fog droplet, as $K = (\varepsilon - 1)/(\varepsilon + 2)$ and the complex permittivity of liquid water $\varepsilon$ is modeled using the double Debye dielectric model (D3M) [19]. In this work, filtered water was used for fog generation, and thus dielectric parameters of pure water are employed. At each second, the measured droplet histogram from the laser particle size analyzer is fitted to the modified gamma distribution, and Eq. (1) is used to compute time-resolved power loss.

In addition to droplet-induced extinction (power loss), the microphysical process of fog creation inevitably increases the ambient humidity within the propagation environment. Therefore, the total channel attenuation must account for the degradation caused by water vapor absorption in addition to droplet-induced extinction. This process is described as the sum of discrete spectral absorption lines of water vapor and a broadband continuum component. A line-by-line calculation ($\gamma_m$) including the continuum is adopted based on the MPM93 model following the ITU-R P.676-13 method [20, 21]. This approach has been validated up to 450 GHz, though its accuracy decreases at higher frequencies [22]. The dynamic model is therefore the comprehensive superposition of the DSD-driven extinction and the humidity-driven molecular absorption, as

$$\gamma_{total}(t) = \gamma_f(t) + \gamma_m(t) \qquad (4)$$

with $\gamma_m(t)$ representing the humidity-driven molecular absorption based on the ITU-R P.676-13 method.

## 3. Channel measurement and modeling

Reference measurements without fog were first conducted to establish baseline power levels and confirm system linearity and stability. These calibration runs provided the foundation for subsequent fog experiments by ensuring that both the THz channel power readings and microphysical observations were synchronized on a common time grid. Then, fog was introduced into the chamber, and simultaneous measurements of received power were acquired together with DSD data from the particle size analyzer. Ambient chamber conditions were continuously monitored, with an initial room temperature of 24.8 °C and humidity of 44.6% RH. During fog trials, humidity levels increased sharply, allowing the contribution of water vapor absorption to be quantified alongside droplet-induced losses. To guarantee reliable microphysical retrieval, fog output was carefully regulated so that the laser beam of the analyzer could penetrate the chamber even under dense conditions.

Two sets of experiments were performed, one at 220 GHz and one at 320 GHz. Measurement results are shown in Fig. 2 (a) and (b). In both cases, the measured power loss rises sharply during the initial 30 s as the chamber fills, before stabilizing into a slower drift with superimposed short-term fluctuations. The model by Eq. (4) closely reproduces the steady-state regime, but underestimates the rapid onset phase. This discrepancy arises from limitations of the particle size analyzer, which cannot fully capture the fast variation of droplet number concentration $N_0$ in the confined chamber volume. Such limitations are aggravated by the small chamber dimensions, where fog density builds too quickly for reliable microphysical sampling. To address this limitation, subsequent experiments employ a larger room-scale chamber to moderate the initial transients.

It is worth emphasizing that the measured 320 GHz channel consistently exhibits greater power loss than the 220 GHz channel. This observation agrees with the spectral prediction in Fig. 2(c), obtained using retrieved distribution parameter $\alpha$=9.98, $\beta$=0.2452, and $b$=1.9731, corresponding to the fog state at 180 s in Fig. 2(b) under 24.4 °C and 92.2% RH (Fig. 2(b)). The frequency-dependent trend highlights the distinction between fog and rain. Whereas in rain both absorption and scattering contribute significantly to total power loss [5], in fog the small droplet diameter relative to the THz wavelength suppress scattering effects. As a result, absorption dominates, making it the primary mechanism of fog-induced power loss in the studied frequency range. This is consistent with the negligible of scintillation effect suffered by the THz channel, as indicated in reference [23], and further justifies the use of Rayleigh approximation models for practical channel prediction. This means that performance degradation in THz fog channels is largely deterministic and spectrally dependent, unlike the highly stochastic fading experienced in optical links through fog. Thus, adaptive system design can focus on frequency-aware link budgeting and power control, rather than on complex diversity or scintillation-mitigation schemes.

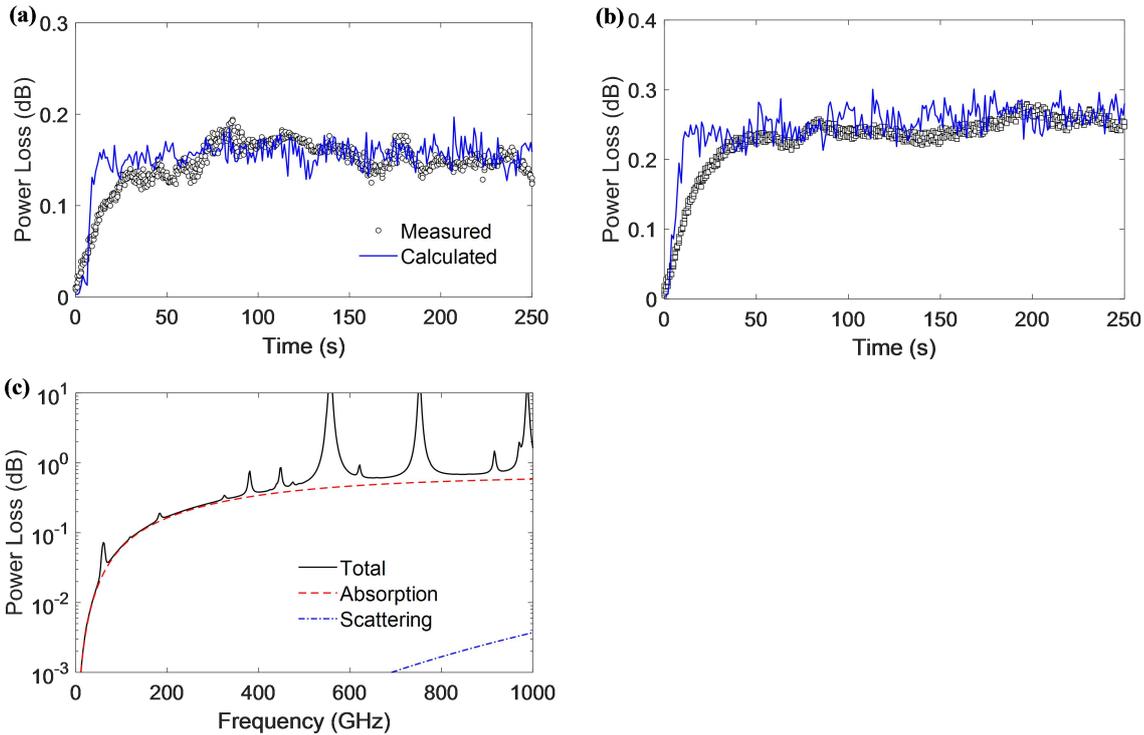

**Figure 2** Measured and modeled temporal attenuation at (a) 220 GHz and (b) 320 GHz, including (c) spectral power loss prediction. (b) keeps the same legend with (a).

## 4. Channel power profile

To assess the small-scale statistics of the received signal, the cumulative distribution function (CDF) of the measured SNR was analyzed during the steady-state phase of the fog experiments, as in Fig. 2(a) and (b). Empirical CDFs were compared against two common amplitude models - the Rician and Weibull distributions. These models were chosen because, under absorption-dominated conditions, scattering is negligible, making Rician a plausible description of the line-of-sight (LoS)-dominated channel [24].

As shown in Fig. 3, the Rician distribution provides an excellent fit to the measured SNR statistics across the entire probability range, outperforming the Weibull model. The fitted K-factors are consistently high - approximately 41.4 dB at 220 GHz and 38.1 dB at 320 GHz (see Table I) - indicating a strong LoS component with only limited diffuse contributions [19]. This behavior implies that even in dense fog, THz channels maintain reliability comparable to clear-air conditions, as the dominant LoS path mitigates scattering-induced fading. The modest differences in K-factor across frequencies (over a 100 GHz span from 220 GHz to 320 GHz) reflect only a weak frequency dependence of the LoS-to-diffuse balance, with LoS stability being the governing factor. This observation suggests that outage and margin calculations can reliably adopt a Rician fading model with high K-factor, while temporal SNR fluctuations should be modeled through the evolving DSD rather than by altering small-scale fading assumptions [5, 25].

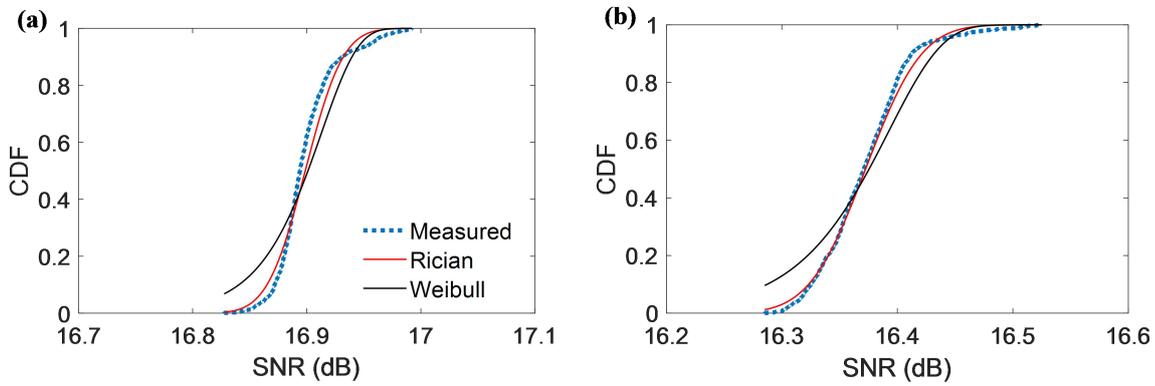

**Figure 3** Cumulative distribution function (CDF) of received SNR of (a) 220 GHz and (b) 320 GHz channels in fog, fitted with Rician and Weibull models. (b) keeps the same legend with (a).

Table I Fit metrics for Rician and Weibull distributions

| Frequency (GHz) | Channel | $R^2$ | Fitting Gap | Fluctuation Error | K-factor (dB) | Shape parameter | Scale parameter |
|---|---|---|---|---|---|---|---|
| 220 | Rician | 0.9879 | 0.034455 | 0.027754 | 41.408 | | |
| | Weibull | 0.94464 | 0.077854 | 0.053806 | | 92.586 | 43.578 |
| 320 | Rician | 0.99702 | 0.017032 | 0.011841 | 38.071 | | |
| | Weibull | 0.96304 | 0.056773 | 0.045934 | | 135.21 | 49.123 |

## III. Channel performance in fogy room

**1. Channel sounding system architecture**

As discussed earlier, the confined volume of the fog chamber leads to excessively rapid changes in droplet number concentration during fog onset, outpacing the response of the particle size analyzer and creating discrepancies between modeled and measured attenuation. To overcome this limitation, we extended the study to a room-scale fog environment, which promotes more gradual fog evolution and complex mixing dynamics, better approximating the quasi-non-stationary nature of natural fog. In this configuration, efficient mixing produced near-homogeneous fog distributions at the meter scale, enabling the retrieved DSD to serve as a reliable representation of the actual conditions along the channel path.

The room-scale experiments were conducted in an enclosed space measuring 5.5×4.0×4.0 m$^3$ (L×H×W), with a 3.3 m line-of-sight (LoS) THz channel path, as shown in Fig. 4. Both transmitter and receiver employed high-gain horn antennas coupled with 15 cm dielectric lenses, calibrated to achieve directional gains up to 40 dBi at 220 GHz. The antennas were mounted 135 cm above the floor to suppress ground reflections and maintain channel stability. Dense fog was generated using a different commercial ultrasonic fog generator with a rated capacity of 30 kg/h, ensuring sustained droplet concentrations representative of atmospheric fog, with diameters ranging from 1 to 40 μm. Microphysical properties were characterized using the same laser particle size analyzer (Furbs FBS-310B ) as in the chamber tests. To avoid interference with the THz beam, the laser probe was again positioned at an angle rather than collinear with the transmission path. In the larger volume, efficient mixing produced near-homogeneous fog distributions at the meter scale, allowing the retrieved DSD to serve as a reliable approximation of the actual conditions along the THz channel path. Measurements were performed at both 220 and 320 GHz using the same instrumentation chain and calibration procedures as in the chamber experiments. Once the THz system was activated, fog was introduced until uniformity was achieved, and time-varying DSDs were recorded, confirming the dynamic nature of droplet distributions during the trials (see Fig. 4 inset).

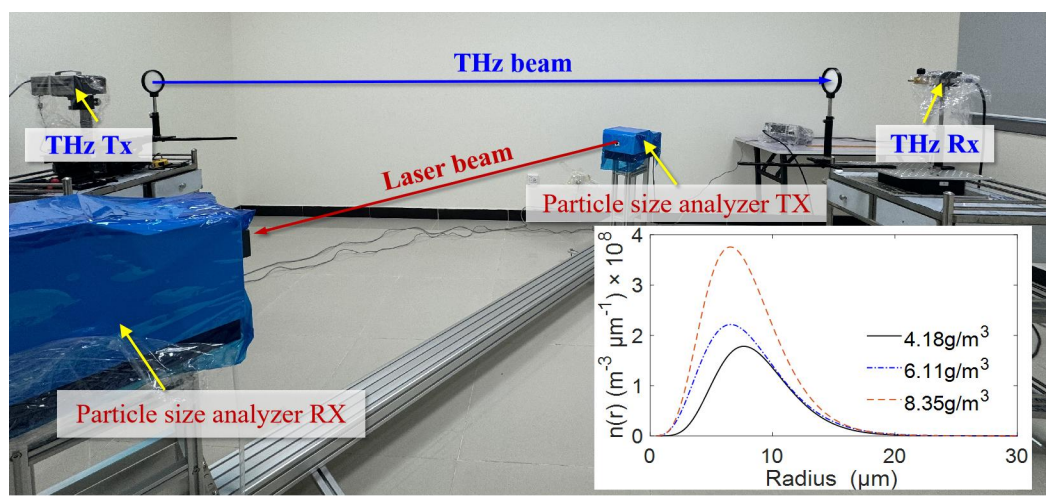

**Figure 4** Experimental setup for channel measurement in fogy room. Inset: measured DSD variation during fog evolution.

## 2. Channel measurement and modeling

The room-scale experiments revealed important differences compared with the controlled fog chamber tests. In the fog chamber, environmental parameters could be precisely regulated and maintained, whereas in the larger room the injection of fog resulted in less control over density and spatial uniformity. As shown in Fig. 5, the fog-induced power loss increased steadily with time, and even after 400 s a stable equilibrium was not reached. This reflects the slower dynamics of fog buildup in the larger space, which better resembles real atmospheric conditions but complicates precise reproducibility.

Modeling predictions by Eq.(4) were compared with the measured data. Agreement was satisfactory only during intermediate stages of the experiment (highlighted in red in Fig. 5), while significant discrepancies appeared at both the onset and saturation phases. At the beginning of fog release, the optical obscuration of the analyzer's laser beam was extremely low, approaching the instrument's sensitivity threshold. Under these conditions, variations in transmitted laser signal were insufficient to distinguish fog from background noise, leading to underestimated droplet counts. Conversely, during the late stage of the experiment, the fog became so dense that the laser beam was strongly attenuated and failed to fully penetrate through the fog. This drove the analyzer into a saturation regime, reducing received optical power and causing systematic underestimation of droplet concentrations and distortion of the inferred DSD.

Within the validated retrieval windows the Rayleigh-based attenuation model, combined with the water vapor absorption term, reproduced both the drift and fluctuations of the measured power loss. Deviations observed outside the retrieval range were consistent with the identified measurement system constraints. Together, the data and predictions confirm that dynamic DSD-driven modeling is essential for accurate description of THz channel propagation in dense fog. The temporal evolution of the DSD governs both the absolute loss level and short-term variability, which cannot be captured by static proxies such as fixed LWC or visibility [14, 26]. It should be noted that we did not analyze the channel power profile in the foggy room because, unlike the chamber where a quasi-stationary state is reached, the fog density in the larger room kept evolving throughout the measurement, making the channel inherently non-stationary and unsuitable for statistical fading characterization [27].

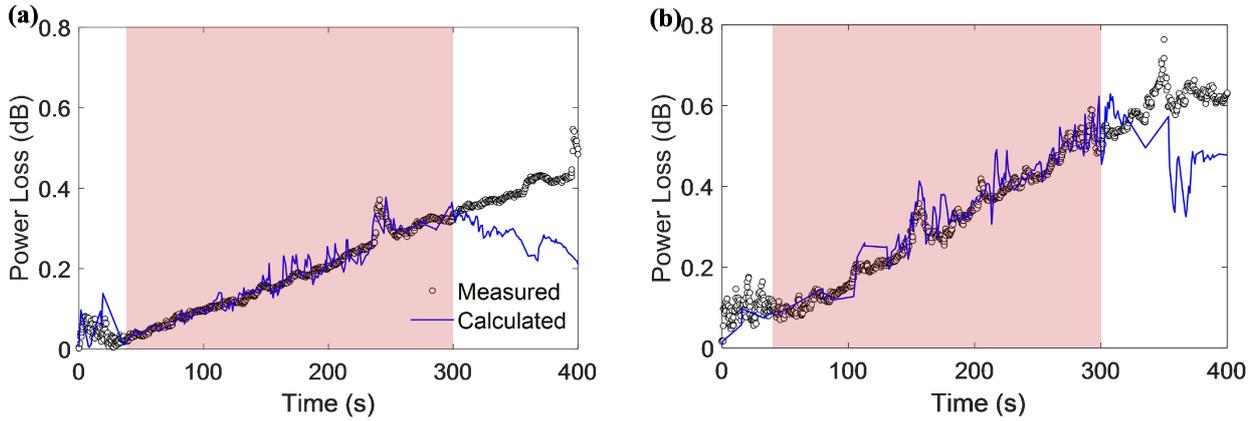

**Figure 5** Measured and modeled temporal attenuation at (a) 220 GHz and (b) 320 GHz, with valid inversion windows highlighted.(b) keeps the same legend with (a).

To assess the impact of the scattering formulation on attenuation retrievals, we computed the fog-induced power loss at 220 and 320 GHz using both the full Mie solution and its Rayleigh approximation, driven sample-by-sample by the DSDs retrieved within the validated obscuration window. Fig. 6 compares the two predictions using matched time stamps. Across the entire observed loss range, the points lie nearly on the 1:1 line, indicating practical equivalence of the two models under our experimental conditions.

The near-equivalence between Rayleigh and Mie predictions can be directly attributed to the microphysics‐wavelength scaling. For fog droplets with diameter $r \sim$ 1–40 μm and millimeter wavelengths at 220 GHz and 320 GHz, the size parameter $\chi=2\pi r/\lambda$ remains well below unity. In this regime, the full Mie solution asymptotically reduces to the Rayleigh approximation, and the scattering cross-section is several orders of magnitude smaller than the absorption component [9, 28, 29]. As a result, fog-induced powerloss is absorption-dominated and scales linearly with the first moment of the DSD. This conclusion is reinforced by the spectral predictions in Fig. 6(c), which show that Rayleigh- and Mie-based calculations remain nearly identical up to about 1 THz, diverging only at higher frequencies where scattering begins to contribute. Therefore, for operating frequencies below 1 THz covering most envisioned THz communication bands - Rayleigh-based models are both accurate and computationally efficient for estimating fog-induced power loss.

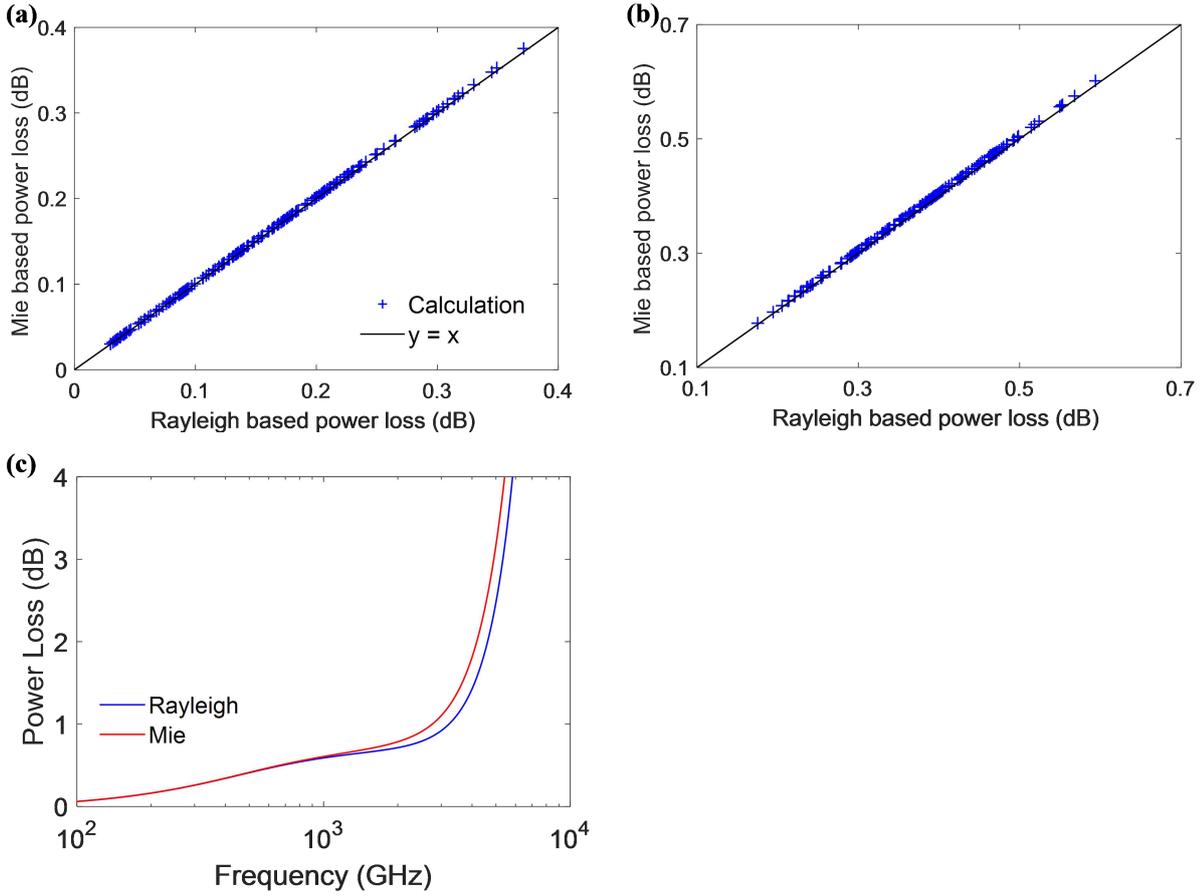

**Figure 6** Comparison of attenuation predictions using Mie theory and Rayleigh approximation under retrieved DSD at (a) 220 GHz and (b) 320 GHz. (c) Predicted power loss by using Rayleigh and Mie scattering theories. (b) keeps the same legend with (a).

## IV. Channel bit error performance

This section translates the previously established physical and statistical characteristics of THz channels under fog into system-level performance implications by analyzing bit error rate (BER). A central design question is whether fog-induced degradation arises primarily from rapid multipath fading or from slower variations in signal-to-noise ratio (SNR). While previous publications of THz channels in fog have largely emphasized power loss and phase perturbations [7, 9], the system-level impacts, especially on BER quantitatively, remain insufficiently explored. By extending the modeling framework, this section evaluates BER performance under representative conditions and modulation schemes, providing a quantitative basis for understanding how droplet size distribution (DSD)-driven attenuation governs channel reliability.

Simulations of a 150 m channel, parameterized by the Rician fading characteristics obtained from our indoor measurements (Section II), allow comparison of QPSK and 16-QAM modulation. QPSK encodes information via phase changes and is inherently more robust to amplitude fluctuations [30, 31], whereas 16-QAM provides higher spectral efficiency at the cost of greater sensitivity to amplitude distortion [32, 33]. Under conventional AWGN channels, the BER expressions of both schemes follow well-known analytical forms [34], as

$$\text{BER}_{QPSK} = \frac{1}{2}\text{erfc}\sqrt{\frac{\gamma}{2}} \quad (5)$$

$$\text{BER}_{16QAM} = \frac{3}{8}\text{erfc}\sqrt{\frac{\gamma}{10}} \quad (6)$$

where $\gamma$ represents the SNR. In fog, however, the SNR distribution deviates from Gaussian statistics; the average BER is obtained by integrating the conditional BER over the probability density function (PDF) of the SNR under the Rician distribution [35, 36], as

$$\text{BER}_{QPSK} = \int_0^\infty \frac{1}{2}\text{erfc}\sqrt{\frac{\gamma}{2}} \cdot f(\gamma) d\gamma \quad (7)$$

$$\text{BER}_{16QAM} = \int_0^\infty \frac{3}{8}\text{erfc}\sqrt{\frac{\gamma}{10}} \cdot f(\gamma) d\gamma \quad (8)$$

For a Rician fading channel, the PDF of the instantaneous symbol SNR $\gamma$ is given by

$$f(\gamma) = \frac{1+K}{\bar{\gamma}}\exp\left(-K - \frac{(1+K)\gamma}{\bar{\gamma}}\right) I_0\left(2\sqrt{\frac{K(1+K)\gamma}{\bar{\gamma}}}\right) \quad (9)$$

with $\bar{\gamma}$ being the average SNR and $I_0(\cdot)$ as the modified Bessel function of the first kind and order zero. All relevant channel parameters, including transmit power, antenna gains, lens gains, and noise levels, are summarized in Table II. Here we choose a finxed of K-facor 30 dB for the Rician fading, as the continuous evolution fo the received channel power, which makes the recording of K-factor impossible. These values are smallers than that for the conditions in fog chamber.

Table II Parameters for channel bit error modeling

| Parameter | Value |
|---|---|
| Transmit power | 10 dBm |
| Frequency | 220 GHz/320GHz |
| Channel length | 150 m |
| Transmitting/Receiving antenna gain | 25 dBi |
| Transmitting/Receiving lens gain | 20 dBi |
| Noise power | -50 dBm |
| Channel distance | 150 m |

Figure 7(a) presents the time-resolved BER evolution at 220 GHz under both AWGN and Rician fading conditions during fog formation. The two curves are nearly identical, showing that BER follows the same trajectory regardless of the fading model. This overlap is attributed to the exceptionally high K-factors observed in our measurements, which exceed 30 dB and confirm the overwhelming dominance of the LOS path. Under such conditions, the scattered component contributes negligibly, and the Rician distribution effectively reduces to an AWGN channel model. This finding is consistent with the absorption-dominated nature of fog-induced attenuation, as previously demonstrated in Fig. 2(c). Hence, the BER dynamics are not driven by multipath fading but by slow variations in power loss directly linked to the evolving DSD of fog. This implies that THz channelss in fog can be treated as near-Gaussian channels with time-varying SNR, simplifying performance prediction and system design.

The temporal structure of BER at both 220 and 320 GHz, shown in Fig. 7(a) and (b), further supports this interpretation. In both frequency bands, BER exhibits a two-stage evolution - an abrupt increase during the initial onset of fog, followed by slower drifts and fluctuations as the fog field matures. This trend mirrors the time-resolved power loss in Fig. 5, highlighting the direct influence of fog microphysics. During the early stage, the rapid rise in droplet concentration causes a steep drop in SNR, leading to a sudden jump in BER. Once the fog distribution becomes more homogeneous, the DSD evolves more gradually, producing smoother SNR variations that manifest as moderate BER drifts with superimposed second-scale fluctuations. These observations imply that BER performance in fog is dictated not by fast fading but by gradual changes in attenuation tied to DSD evolution. For future system design, this reinforces the need for adaptive modulation and coding that tracks time-varying SNR [37, 38], rather than diversity schemes aimed at countering rapid multipath fading.

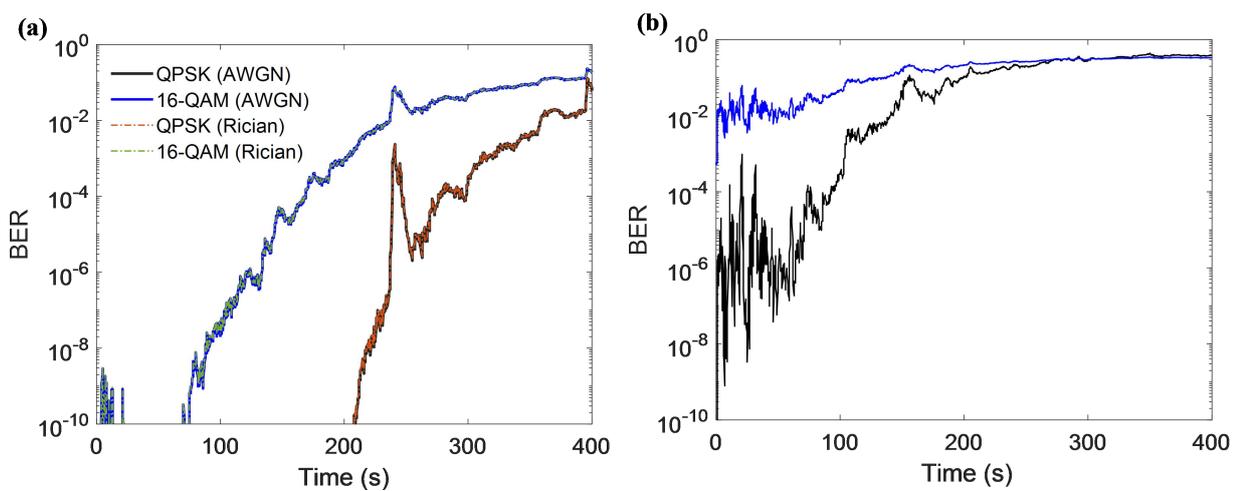

**Figure 7** Bit error rate (BER) evolution under fog conditions for QPSK and 16-QAM modulation. (a) Comparison of AWGN and Rician fading channel models (b) BER performance under AWGN channel model. (b) keeps the same legend with (a).

## V. Conclusion

This article presented an experimental characterization and dynamic modeling of THz channel performance at 220 GHz and 320 GHz under controlled and quasi-natural fog environments. The integration of real-time microphysical sensing with electromagnetic modeling leads to several key findings. The microphysics-aware DSD model accurately captures temporal channel power dynamics and confirms that extinction is overwhelmingly absorption-dominated at frequencies below 1 THz, providing strong justification for the use of the computationally efficient Rayleigh approximation in practical link budget calculations. Statistical analysis of SNR confirmed exceptional channel stability, characterized by high Rician K-factors, which proves that LoS dominance is maintained even in dense fog due to minimal scattering. BER analysis demonstrated that performance degradation is governed entirely by the slow DSD evolution (time-varying SNR) rather than fast multipath fading. Consequently, the THz fog channel can be reliably modeled as a near-Gaussian channel with time-varying SNR, redirecting adaptive strategies toward SNR tracking and frequency-aware link budgeting, rather than complex spatial diversity.

Our future work should extend these investigations to outdoor environments with natural fog and longer distances, while also accounting for the spatial heterogeneity of fog to better characterize and model its impact on THz channel performance for practical 6G applications.